\documentclass[prl,showpacs,twocolumn]{revtex4}%
\usepackage{graphicx}
\usepackage{bm}
\usepackage{amsmath}
\usepackage{amsfonts}
\usepackage{amssymb}%
\providecommand{\U}[1]{\protect\rule{.1in}{.1in}}
\begin{document}
\title{Intrinsic spin Hall effect in graphene: Numerical calculations in a
multi-orbital model}
\author{Seiichiro Onari}
\author{Yasuhito Ishikawa}
\author{Hiroshi Kontani$^{1}$}
\author{Jun-ichiro Inoue}
\affiliation{Department of Applied Physics, Nagoya University, Chikusa, Nagoya 464-8603, Japan}
\affiliation{$^{1}$Department of Physics, Nagoya University, Chikusa, Nagoya 464-8603, Japan}
\date{\today}

\begin{abstract}
We study the spin Hall effect (SHE) in graphene using a realistic
multi-orbital tight-binding model that includes the atomic spin-orbit
interaction. The SHE is found to be induced by the spin-dependent
Aharonov-Bohm phase. In the metallic case, the calculated values for the spin
Hall conductivity (SHC) are much smaller than the quantized Hall conductivity
for realistic parameter values of metallic graphene. In the insulating case,
quantization of the SHC is violated due to the multi-orbital effect. The
present study suggests that the SHE in a honeycomb lattice is enhanced by
chemical doping, such as the substitution of carbon atoms with boron atoms.

\end{abstract}

\maketitle



The recent discovery of graphene has stimulated much interest in electron
transport in graphene due to its unconventional electronic
structure.\cite{novoselov1} The crystal structure of graphene is a
two-dimensional honeycomb lattice of carbon atoms. In the absence of the
spin-orbit interaction (SOI), the band structure of graphene is described in
terms of massless Dirac fermions at the edges (K and K' points) of the
Brillouin zone. Due to a non-zero Berry's phase (a geometric quantum
phase)\cite{ando-berry}, a striking \textquotedblleft
half-integer\textquotedblright\ quantum Hall effect is realized in
graphene\cite{novoselov2,zhang-graphene}. The nature of the quantum Hall
effect near the massless Dirac point has been studied
theoretically.\cite{gusynin,abanin,nomura,sheng,ezawa,hatsugai-aoki}
Concerning transport phenomena in graphene, the emergence of a sizable spin
Hall effect (SHE), which is a phenomenon that a spin current flows
perpendicular to an applied electric field, had been predicted
theoretically.\cite{kane,sinitsyn} In particular, Kane and Mele\cite{kane}
have demonstrated that the SOI generates an energy gap at the Dirac point, and
the ``quantum SHE" is expected to appear in insulating graphene.

Recently, the theory of the intrinsic SHE caused by the Berry phase in
zincblend semiconductors\cite{murakami-SHC} or by a uniform SOI such as the
Rashba SOI in a two-dimensional electron gas\cite{sinova} has been attracting
much interest. Quite recently, the SHE has also
been observed in the simple metal Al\cite{valenzuela} and the transition metal
Pt.\cite{kimura-SHE} The spin Hall conductivity (SHC) reported for Pt is
$240(\hbar/e)(\Omega\mathrm{cm})^{-1}$ at room temperature, which is much
larger than that of semiconductors. The huge SHC in Pt has been explained
theoretically by Kontani \textit{et al.} in terms of the atomic SOI and a
realistic tight-binding (TB) model that includes inter-orbital hopping
integrals.\cite{kontani-SRO,kontani-pt,tanaka-kontani} They further showed
that the so-called current vertex corrections (CVC) have little effect on the
SHC in transition metal, in contrast to that in semiconductors with the Rashba
SOI.\cite{inoue}

The successful reproduction the experimental values of the SHC in transition
metals by the TB model with the atomic SOI encouraged us to apply this model
to other systems. Since the electronic structure of graphene is well
reproduced by the realistic TB model, we use this model to calculate the SHCs
in both undoped and doped (metallic) graphene, and compare the results with
those calculated using the Kane-Mele model,\cite{kane} which contains only a
$p_{z}$ orbital and an effective SOI to produce an energy gap at the Fermi energy.

In the present paper, we adopt a realistic multi-orbital ($s$, $p_{x}$,
$p_{y}$ and $p_{z}$ orbitals) TB model with the atomic SOI on a honeycomb
lattice, and use the Kubo-Streda formula\cite{streda} to calculate the SHC. It
should be noted that the $p_{x}$ and $p_{y}$ orbitals contribute to the SHC
because they are mixed with the $p_{z}$ orbital via the atomic SOI. By
calculating the SHC as a function of the Fermi energy, we find that the SHC
becomes large in the energy region where the $p_{x}$ and $p_{y}$ orbitals are
dominant, far away from the Dirac point. The effect of the CVC, which is
calculated in the self-consistent Born approximation, is appreciable and
causes the SHC to double compared to that without the CVC, while the
qualitative behavior of the SHC remains unchanged. In the case of insulating
graphene, SHC is not quantized because $s_{z}$ is not conserved due to the SOI
between the $p_{x}(p_{y})$ and $p_{z}$ orbitals.

The Hamiltonian for electrons on the honeycomb lattice, which is decomposed
into A and B sub-lattices, is given as $\hat{H}=\hat{H}_{0}+\hat
{H}_{\mathrm{SO}}$ where $\hat{H}_{0}$ and $\hat{H}_{\mathrm{SO}}$ are the
kinetic and SOI terms, respectively. $\hat{H}_{0}$ is given as
\begin{equation}
\hat{H}_{0}=\sum_{i\subset\mathrm{A(B)}}\sum_{j\subset\mathrm{B(A)}}%
\sum_{\alpha\beta\sigma}t_{ij}^{\alpha\beta}c_{i\sigma}^{\alpha\dagger
}c_{j\sigma}^{\beta},
\end{equation}
where $t_{ij}^{\alpha\beta}$ denotes the nearest-neighbor hopping integral
between the $\alpha$ orbital at site $i$ and the $\beta$ orbital at site $j $,
and $\sigma$ denotes the spin.
The Slater-Koster parameters in the TB model are taken as $ss\sigma=-0.86$,
$sp\sigma=1.14$, $pp\sigma=1.0$ and $pp\pi=-0.5$ in units of $pp\sigma
(5.7$eV).\cite{Harrison} Hereafter, we take $pp\sigma$ and the lattice
constant $a(2.55$\AA ) as the units for energy and length, respectively. We
also put $\hbar=1$.

$\hat{H}_{\mathrm{SO}}$ represents the atomic SOI given as
\begin{equation}
\hat{H}_{\mathrm{SO}}=\lambda\sum_{i}\bm{l}_{i}\cdot\bm{s}_{i},
\end{equation}
where $\lambda$ is the interaction constant, and $\bm{l}$ and $\bm{s}$ are the
orbital and spin angular momentum, respectively. $\hat{H}_{\mathrm{SO}}$ has
off-diagonal elements such as $\langle p_{x}\pm|\hat{H}_{\mathrm{SO}}|p_{y}%
\pm\rangle=\mp i\lambda/2$, $\langle p_{x}\pm|\hat{H}_{\mathrm{SO}}|p_{z}%
\mp\rangle=\pm\lambda/2$ and $\langle p_{y}\pm|\hat{H}_{\mathrm{SO}}|p_{z}%
\mp\rangle=-i\lambda/2$, where $|l\sigma\rangle$ represents an electron with
orbital $l$ and spin $\sigma=\pm1$.

According to Streda,\cite{streda} the intrinsic SHC at $T=0$ is given as
$\sigma_{xy}^{z}=\sigma_{xy}^{z\mathrm{I}}+\sigma_{xy}^{z\mathrm{II}}$, where
\begin{equation}
\sigma_{xy}^{z\mathrm{I}}(E)=\frac{1}{2\pi N}\sum_{\bm{k}}\mathrm{Tr}\left[
\hat{J}_{x}^{\mathrm{S}}\hat{G}^{\mathrm{R}}\hat{J}_{y}^{\mathrm{C}}\hat{G}^{
\mathrm{A}}\right]  _{\omega=E}, \label{sigmaI}%
\end{equation}
\begin{align}
\sigma_{xy}^{z\mathrm{II}}(E)  &  =\frac{-1}{4\pi N}\sum_{\bm{k} }%
\int_{-\infty}^{E}d\omega\mathrm{Tr}\left[  \hat{J}_{x}^{\mathrm{S}}\frac{
\partial\hat{G}^{\mathrm{R}}}{\partial\omega}\hat{J}_{y}^{\mathrm{C}}\hat{
G}^{\mathrm{R}}\right. \nonumber\\
&  -\left.  \hat{J}_{x}^{\mathrm{S}}\hat{G}^{\mathrm{R}}\hat{J}_{y}%
^{\mathrm{C} }\frac{\partial\hat{G}^{\mathrm{R}}}{\partial\omega}%
-\langle\mathrm{R} \leftrightarrow\mathrm{A}\rangle\right]  . \label{sigmaII}%
\end{align}
Here, $\hat{G}^{\mathrm{R(A)}}$ is the retarded (advanced) Green's function
given as $\hat{G}^{\mathrm{R(A)}}(\bm{k},\omega)=\left(  \omega-\hat
{H}+(-)i\hat{\Gamma}\right)  ^{-1}$, where $\hat{\Gamma}$ is the matrix form
of the imaginary part of the self-energy (damping rate) due to scattering by
local impurities. The Green's functions are represented by $16\times16$
matrices in the momentum representation.

The matrix forms of the charge and $s_{z}$-spin current operators for the
$\mu$-direction $(\mu=x,y)$ are given by $\hat{J}_{\mu}^{\mathrm{C}}=-e\frac{
\partial\hat{H}}{\partial k_{\mu}}$ and $\hat{J}_{\mu}^{\mathrm{S}}=- \frac
{1}{2e}\left\{  \hat{J}_{\mu}^{\mathrm{C}},s_{z}\right\}  $, respectively. In
the SHC, $\sigma_{xy}^{z\mathrm{I}}$ and $\sigma_{xy}^{z \mathrm{II}}$
represent the \textquotedblleft Fermi surface term\textquotedblright\ and the
\textquotedblleft Fermi sea term\textquotedblright, respectively. In this
paper, we take $N=900\times900 $ $\bm{k}$-point meshes in the numerical calculations.

\begin{figure}[pth]
\begin{center}
\includegraphics[width=70mm]{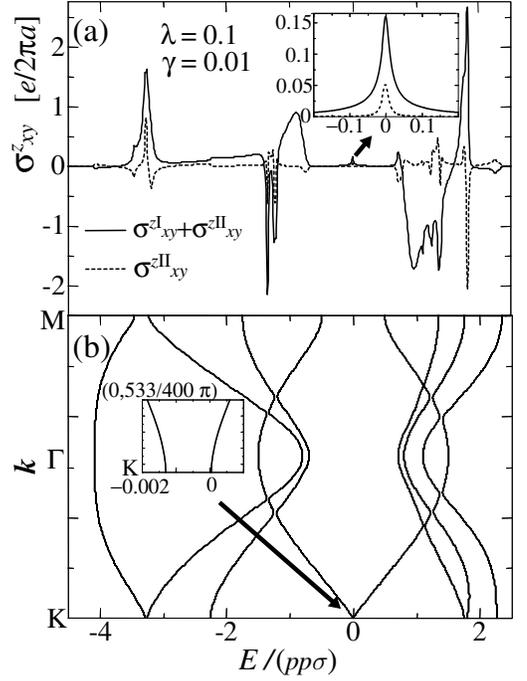}
\end{center}
\caption{(a)SHC $\sigma_{xy}^{z\mathrm{I}}+\sigma_{xy}^{z\mathrm{II}}$ (solid
line) and $\sigma_{xy}^{z\mathrm{II}}$ (dotted line), and (b) band structure
through $\bm{k}$ points K$=(0,4\pi/3)$ , $\Gamma=(0,0)$ and M$=(\pi/\sqrt
{3},\pi)$ against energy for atomic SOI $\lambda=0.1$ and damping
$\gamma=0.01$ (in units of $pp\sigma$). The insets show the SHC and band
structure near the Dirac point.}%
\label{LS-sigma-Disp}%
\end{figure}

We first neglect the CVC and assume that the damping matrix is diagonal and is
independent of orbital $\hat{\Gamma}_{\alpha\beta}=\gamma\delta_{\alpha\beta}%
$. Figures \ref{LS-sigma-Disp} (a) and (b) give the calculated results of the
SHC as a function of the position of the Fermi energy $E_{F}$ and the band
structure of graphene for $\lambda=0.1$ and $\gamma=0.01$ (in units of
$pp\sigma$), respectively. In Fig. \ref{LS-sigma-Disp}(a), the solid and
dotted lines denote $\sigma_{xy}^{z}$ and $\sigma_{xy}^{z\mathrm{II}}$,
respectively. We observe that the SHC $\sigma_{xy}^{z}$ is large in energy
regions where the bands are nearly degenerate, while $\sigma_{xy}%
^{z\mathrm{II}}$ exhibits peak structures in these energy regions. In the
metallic region, the magnitude of $\sigma_{xy}^{z\mathrm{I}} $ is much larger
than that of $\sigma_{xy}^{z\mathrm{II}}$.

The insets of Figs. \ref{LS-sigma-Disp}(a) and (b) show the SHC and the band
structure near the Dirac point of graphene, i.e., $E_{F}=0$. The Dirac point
at K is split by the atomic SOI, and forms an energy gap with $\Delta
_{\mathrm{D}}=0.0013$ when $\lambda=0.1$. 

In order to study the SHE in graphene, we estimate the value of $\gamma$ to be
about $0.005$ by comparing the longitudinal conductivity $\sigma_{xx}%
$\ calculated in the present model with the experimental resistivity observed
for doped graphene $\rho=1/\sigma_{xx}=100\Omega$.\cite{zhang} In this case,
the number of electrons $n$ per atom of doped graphene is $n=3.998$, which
corresponds to $E_{F}=-0.05$ in the present model. The value of $\lambda$ of
the atomic SOI at the Dirac point in graphene is estimated to be
$\lambda=0.001$,\cite{min,hernando} which corresponds to an energy gap
$\Delta_{\mathrm{D}}=2\times10^{-7}(\sim10^{-6}\mathrm{eV})$. The SHC
calculated by using $\lambda=0.001$ and $\gamma=0.005$, and by neglecting the
energy dependence of $\gamma$ is $\sigma_{xy}^{z}\sim3\times10^{-5}[e/2\pi a]$
at the Dirac point. This value is much smaller than the quantized SHC $e/2\pi
a$, since $\gamma\gg\Delta_{\mathrm{D}}$ and the graphene is metallic for
these parameter values.

The SHC may increase in doped graphene; for example, the value of $\sigma
_{xy}^{z}$ increases to $\sim0.1[e/2\pi a]$ at $E_{F}=-0.76$, which can be
realized by substituting $50\%$ carbon atoms with boron atoms. Although this
situation may be virtual, similar electronic state may be realized by graphite
intercalation. Since $1[e/2\pi a]\sim1500(\hbar/e)(\Omega\mathrm{cm} )^{-1}$
in the present case ($a=2.55$\AA ), the SHC $\sim150(\hbar/e)(\Omega
\mathrm{cm})^{-1}$ at $E_{F}=-0.76$ is the same order as that of Pt.

\begin{figure}[pth]
\begin{center}
\includegraphics[width=60mm]{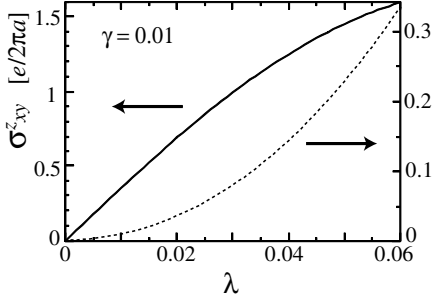}
\end{center}
\caption{$\sigma_{xy}^{z}$ against $\lambda$ with $\gamma=0.01$ (in units of
$pp\sigma$) for $n=3.2$ $(E_{F}=-1)$ (solid line) and for $n=4$ $(E_{F}=0)$
(dotted line), respectively.}%
\label{lambda-depend2}%
\end{figure}

\begin{figure}[pth]
\begin{center}
\includegraphics[width=70mm]{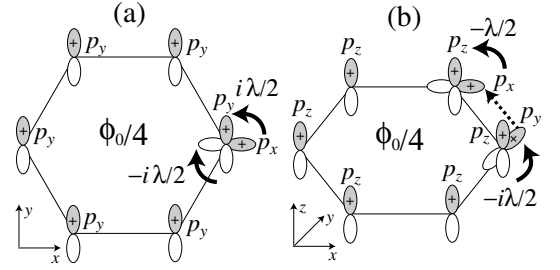}
\end{center}
\caption{Effective AB phase in honeycomb lattice derived by a motion of an
up-spin electron (a) on mainly $p_{x}$ and $p_{y}$ orbitals, and (b) on mainly
$p_{z}$ orbital.}%
\label{honeycomb}%
\end{figure}

Now, we give an intuitive explanation of why the SHE is induced by the atomic
SOI. The calculated $\lambda$ dependence of the SHC is shown in Fig.
\ref{lambda-depend2}. We see that $\sigma_{xy}^{z}\propto\lambda$ and
$\sigma_{xy}^{z}\propto\lambda^{2}$ for $n=3.2$ $(E_{F}=-1)$ and $n=4$
$(E_{F}=0)$, respectively. The results may be interpreted as follows. Since
the $p_{x}$ and $p_{y}$ orbitals are dominant at $E_{F}\sim-1$, an
anti-clockwise motion of an up-spin electron on $p_{y}$ orbitals in a
honeycomb lattice may be given by Fig. \ref{honeycomb}(a). The arrows in the
figure represent an inter-orbital transition induced by the atomic SOI. In the
process $p_{y}\rightarrow p_{x}(p_{x}\rightarrow p_{y})$, the SOI works once,
and yields a factor $(-)i\lambda/2$, which is the first order of $\lambda$.
The corresponding motion of an electron yields a factor $i=e^{2\pi i/4}$,
which can be interpreted as the Aharonov-Bohm (AB) phase factor $e^{2\pi
i\phi/\phi_{0}}$ $(\phi_{0}=hc/|e|)$, where $\phi$ is effective magnetic flux
$\phi=\oint\bm{A}\cdot d\bm{r}=\phi_{0}/4$ through a honeycomb lattice. Since
the sign of the effective magnetic flux is opposite for down-spin electrons,
up- and down-spin electrons move in opposite directions under an electric
field. By this process, $\sigma_{xy}^{z}\propto\lambda$ is realized in the
region where the $p_{x}$ and $p_{y}$ orbitals are dominant, as indicated by
the solid line in Fig. \ref{lambda-depend2}.

At $E_{F}\sim0$ the $p_{z}$ orbital is dominant, and an important process of an
up-spin electron on $p_{z}$ orbitals may be given by Fig. \ref{honeycomb}(b).
Solid arrows and a dotted arrow represent the inter-orbital transitions
induced by the SOI and by hopping, respectively. The SOI should operate at
least twice while the electron moves around the honeycomb structure, since the
spin is flipped in the transitions $p_{z}\rightarrow p_{y}$ and $p_{x}%
\rightarrow p_{z}$ via the SOI in this case. The corresponding motion yields a
factor of $i=e^{2\pi i/4}$, which corresponds to the AB phase and an effective
magnetic flux. By this process, $\sigma_{xy}^{z}\propto\lambda^{2}$ is
realized in the region where the $p_{z}$ orbital is dominant at the Fermi
level, as indicated by the dotted line in Fig. \ref{lambda-depend2} . Since
$\lambda\ll1$, the SHC near $E_{F}=0$ is about 10 times smaller than that near
$E_{F}=-1$. 

Up until now, the CVC have not been taken into account. However, the CVC may
play an important role for the SHC in graphene, as discussed in Ref.
\cite{sinitsyn} for the Kane-Mele model. In order to study the role of the CVC
in the present four-orbital model, we employ the self-consistent Born
approximation, where orbital-dependent damping $\hat{\Gamma}_{\alpha\beta
}=\gamma_{\alpha}\delta_{\alpha\beta}$ is given by
\begin{equation}
\gamma_{\alpha}(E)=n_{\mathrm{i}}I^{2}\frac{1}{2Ni}\sum_{\bm{k}}\left[
\hat{G}_{\alpha\alpha}^{\mathrm{A}}(\bm{k},E)-\hat{G}_{\alpha\alpha
}^{\mathrm{R}}(\bm{k},E)\right]  ,
\end{equation}
where $n_{\mathrm{i}}$ and $I$ are the density of impurities and impurity
potentials, respectively. The total current ($\widetilde{\hat{J}%
_{y}^{\mathrm{C}}}=\hat{J}_{y}^{\mathrm{C}}+\Delta\hat{J}_{y}^{\mathrm{C}}$)
with the CVC ($\Delta\hat{J}_{y}^{\mathrm{C}}$) is given by the Bethe-Salpeter
equation:
\begin{equation}
\widetilde{\hat{J}_{y}^{\mathrm{C}}}(\bm k,\omega)=\hat{J}_{y}^{\mathrm{C}%
}(\bm k)+\frac{n_{\mathrm{i}}I^{2}}{N}\sum_{\bm k^{\prime}}\hat{G}%
^{\mathrm{R}}(\bm k^{\prime},\omega)\widetilde{\hat{J}_{y}^{\mathrm{C}}}(\bm
k^{\prime},\omega)\hat{G}^{\mathrm{A}}(\bm k^{\prime},\omega),\label{bethe}%
\end{equation}
which is solved self-consistently. Here, we put $n_{\mathrm{i}}I^{2}=0.02$ ,
for which the experimental value of resistivity\cite{zhang} $\rho
=1/\sigma_{xx}=100\Omega$ at $E_{F}=-0.05$ is realized in the present
calculation.
The CVC part for the Fermi surface term is obtained from
\begin{equation}
\Delta\sigma_{xy}^{z\mathrm{I}}(E)=\frac{1}{2\pi N}\sum_{\bm k}\mathrm{Tr}%
\left[  \hat{J}_{x}^{\mathrm{S}}\hat{G}^{\mathrm{R}}\Delta\hat{J}%
_{y}^{\mathrm{C}}\hat{G}^{\mathrm{A}}\right]  _{\omega=E},
\end{equation}
and the total SHC with the CVC is given by $\widetilde{\sigma}_{xy}%
^{z\mathrm{I}}=\sigma_{xy}^{z\mathrm{I}}+\Delta\sigma_{xy}^{z\mathrm{I}}$.

\begin{figure}[pth]
\begin{center}
\includegraphics[width=70mm]{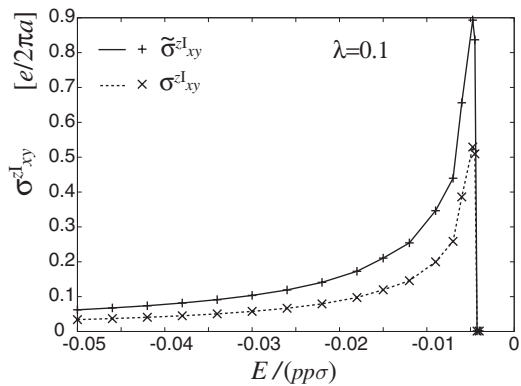}
\end{center}
\caption{Total $\tilde{\sigma}_{xy}^{z\mathrm{I}}$ (solid line) with the CVC
and ${\sigma}_{xy}^{z\mathrm{I}}$ (dotted line) without the CVC for
$\lambda=0.1$ and $\Delta_{\mathrm{D}}=0.0013$.}%
\label{cvc}%
\end{figure}

The calculated results of $\sigma_{xy}^{z\mathrm{I}}$ and $\tilde{\sigma}%
_{xy}^{z\mathrm{I}}$ near the Dirac point are shown in Fig. \ref{cvc} for
$\lambda=0.1$, which makes $\Delta_{\mathrm{D}}=0.0013$ at the Dirac point.
The numerical results for $|E|<0.004$ are omitted because of poor convergence.
We note that $\tilde{\sigma}_{xy}^{z\mathrm{I}}=0$ in the insulating system
$(E_{F}=0)$. We see that $\tilde{\sigma}_{xy}^{z\mathrm{I}} $ with the CVC is
almost double $\sigma_{xy}^{z\mathrm{I}}$ without the CVC. Although the
behavior of $\sigma_{xy}^{z\mathrm{I}}$ is consistent with the results of
Sinitsyn \textit{et al.}\cite{sinitsyn}, $\tilde{\sigma}_{xy}^{z\mathrm{I}}$
is considerably smaller than that obtained by Sinitsyn \textit{et al.} We
consider that the disagreement comes from the difference between two models.
The reason why the CVC remains for the atomic SOI in a honeycomb lattice is
that the Hamiltonian breaks inversion symmetry $\hat{H}_{\alpha\beta
}(\bm{k})\neq\hat{H}_{\alpha\beta}(-\bm{k})$. If $\hat{H}_{\alpha\beta
}(\bm{k})=\hat{H}_{\alpha\beta}(-\bm{k})$ is satisfied, the CVC vanishes
identically. We note that the CVC is absent for the Fermi sea term in the Born
approximation in the present model. 

In the Kane-Mele model, the SHE $\sigma_{xy}^{z}=e/2\pi$ is quantized at the
Dirac point of graphene.\cite{kane} In order to compare with their results, we
calculate $\sigma_{xy}^{z}$ in the energy gap by taking the limit
$\gamma\rightarrow0$, which corresponds to insulating graphene. We obtain
$\sigma_{xy}^{z\mathrm{I}}=0$ and $\sigma_{xy}^{z\mathrm{II}}\sim1.2e/2\pi$ at
$E_{F}=0$ for $\lambda=0.4$. The reason for the violation of the quantization
is that $s_{z}$ is not conserved due to the SOI in the present
model.\cite{murakami-conserve,sheng2}

So far we have neglected the effect of lattice deformation (curvature effect),
which was pointed out to be important for e.g. carbon
nanotube.\cite{ando_curv,hernando, bulaev} The curvature effect induces a
hopping $(\Delta t)$ between $p_{z}$ orbital and $p_{x}(p_{y})$ orbital.
$\Delta t$ is estimated as $\sim0.01$ in graphene\cite{Morozov} and $\sim0.1$
in nanotube\cite{kuemmeth} in units of $pp\sigma$. The lowest order of the
curvature effect on SHC $\sigma_{{xy}\;\mathrm{curv}}^{z}$ is estimated as
$(\Delta t)^{2}\lambda$, where a denominator of order 1eV is omitted. Since
the SHC $\sigma_{{xy}\;\mathrm{int}}^{z}$ without the curvature effect is the
order of $\lambda^{2}$, the ratio $\sigma_{{xy}\;\mathrm{curv}}^{z}%
/\sigma_{{xy}\;\mathrm{int}}^{z}\sim0.1$ in the graphene, and $\sim10$ in the
nanotube by adopting a realistic value of $\lambda=0.001$ for the SOI of
carbon atoms. We expect that the curvature effect for the SHC may be small in
graphene but dominant in nanotube.

In summary, we have studied the SHC in a two-dimensional honeycomb lattice
using a realistic band structure consisting of $s$, $p_{x}$, $p_{y}$ and
$p_{z}$ orbitals with the atomic SOI. The estimated SHC for a realistic value
of the atomic spin-orbit coupling $\lambda$ and constant damping $\gamma$ for the metallic
graphene at the Dirac point is considerably small. We predict that the SHC
will be large when the Fermi level is shifted to $E_{F}=-0.76(\sim
-4.3\mathrm{eV})$, where the $p_{x}$ and $p_{y}$ orbitals are dominant at the
Fermi level.
In the self-consistent Born approximation, the SHC $\tilde{\sigma}%
_{xy}^{z\mathrm{I}}$ with the CVC is almost double the SHC without the CVC.
The obtained value of $\tilde{\sigma}_{xy}^{z\mathrm{I}}$ is considerably
smaller than that of the Kane-Mele model.\cite{kane,sinitsyn} In the case of
insulating graphene, the obtained SHC $\sigma_{xy}^{z\mathrm{II}}$ is not
quantized because $s_{z}$ is not conserved due to the SOI.

This work was supported by a Grant-in-Aid for 21st Century COE
\textquotedblleft Frontiers of Computational Science\textquotedblright\ and a
Grant-in-Aid for Scientific Research in Priority Areas \textquotedblleft
Creation and Control of Spin Current". Numerical calculations were performed
at the supercomputer center, ISSP.

\bibliographystyle{plain}
\bibliography{paper1}

\end{document}